# Multi-Class Plant Leaf Disease Detection: A CNN-based Approach with Mobile App Integration

Md Aziz Hosen Foysal[1], Foyez Ahmed[2], Md Zahurul Haque[3]
Department of Computer Science and Engineering, Manarat International University (MIU), Dhaka, Bangladesh
[1]azizhf9596@gmail.com, [2]ahmedfoyez786@gmail.com, [3]jahurulhaque@manarat.ac.bd

***Abstract*− *Plant diseases significantly impact agricultural productivity, resulting in economic losses and food insecurity. Prompt and accurate detection is crucial for the efficient management and mitigation of plant diseases. This study investigates advanced techniques in plant disease detection, emphasizing the integration of image processing, machine learning, deep learning methods, and mobile technologies. High-resolution images of plant leaves were captured and analyzed using convolutional neural networks (CNNs) to detect symptoms of various diseases, such as blight, mildew, and rust. This study explores 14 classes of plants and diagnoses 26 unique plant diseases. We focus on common diseases affecting various crops. The model was trained on a diverse dataset encompassing multiple crops and disease types, achieving 98.14% accuracy in disease diagnosis. Finally integrated this model into mobile apps for real-time disease diagnosis.***

***Keywords*− *Convolutional Neural Network, Leaf Classification, Disease Diagnosis, Mobile Technologies.***

## I. INTRODUCTION

Plant diseases pose a significant threat to global agricultural productivity, causing substantial financial losses and jeopardizing the health of both humans and animals. Traditional methods of plant disease detection, such as manual inspection, are often time-consuming, less effective, and prone to human error. Farmers, especially in developing nations, rely on these manual methods, which can lead to delays in disease identification and treatment. The integration of Convolutional Neural Networks (CNNs) and artificial intelligence (AI) into plant disease detection offers a promising solution to these limitations. CNNs can process vast amounts of image data quickly and accurately, enabling rapid, precise, and automated identification of plant diseases. This is crucial for timely intervention and mitigation, helping to prevent significant crop losses and maintain food security. This study focuses on leveraging CNNs for plant leaf disease detection and integrating the model into a mobile application for real-time diagnosis, thus providing a practical tool for farmers to manage plant health effectively. Nowadays, technology has greatly improved our lives. With the internet, almost everything is accessible. Using a regular camera, individuals can easily take photos of affected plant parts and upload them to a system that detects the specific disease and suggests the appropriate treatment, including any necessary pesticides. Our study aims to develop accurate models to meet the increasing demand for improved plant disease classification. Leveraging the large dataset, we designed a novel CNN architecture that outperforms several well-known pre-trained and custom models, including Bayesian Optimized SVM + Random Forest, YOLOv5 + Swim Transformer, VGG16, and AlexNet with GAP Layer. The proposed model achieved exceptionally high accuracy, highlighting the importance of large datasets in training robust and reliable disease detection systems. Additionally, the research aims to integrate this high-performing model into a user-friendly Android application using TensorFlow Lite, enabling real-time disease detection through image scanning or uploading. The ultimate goal is to provide farmers with immediate, actionable insights to manage plant health, reduce crop losses, and improve agricultural productivity and sustainability.

## II. RELATED WORKS

Recent developments in plant disease identification have considerably benefitted from the deployment of Convolutional Neural Networks (CNN). CNNs have demonstrated exceptional accuracy in recognizing different plant diseases, making them a preferred choice over traditional methods, which are often time-consuming and less effective. Ferrentinos (2018) provided a comprehensive analysis of deep learning models for plant disease detection, highlighting their ability to diagnose multiple diseases from image data with high precision [3]. This work opened up new avenues for investigation into the application of deep learning to agricultural operations. Building on this foundation, Shrestha et al. (2020) applied CNNs to plant disease detection, demonstrating the robustness of these models in accurately classifying various diseases. Their work, presented at the IEEE Applied Signal Processing Conference, showed promising results and underscored the potential of CNN architectures in enhancing plant pathology diagnostics [2]. Similarly, Deepalakshmi et al. (2021) created a CNN-based method for identifying illnesses in plant leaves, further confirming the effectiveness of CNNs in handling large-scale image data and providing reliable disease classification [4]. Several studies have conducted comparative analyses of different CNN architectures to identify the most effective models for plant disease detection. Sardogan, Tuncer, and Ozen (2018) explored a hybrid approach by combining CNNs with learning vector quantization. (LVQ) for plant leaf disease detection. Their findings indicated that the integration of LVQ with CNNs improved classification accuracy, showcasing the advantages of hybrid models in enhancing diagnostic performance [5]. Agarwal, Gupta, and Biswas (2020) focused on developing an efficient CNN model specifically for tomato crop disease identification.
Their model not only outperformed existing models in terms of accuracy but also demonstrated superior computational efficiency, making it a valuable tool for practical applications in agriculture [7]. The integration of machine learning models into mobile technologies has also been a significant focus in recent research. Wang et al. (2021) developed a trilinear CNN model (T-CNN) for the visual identification of plant illnesses, which was subsequently integrated into a mobile application for real-time diagnosis. This approach allowed farmers to use mobile devices for immediate disease detection, facilitating timely and effective disease management [8]. Joshi and Bhavsar (2023) suggested Night-CNN, a deep learning technology-based system designed mainly for mobile platform deployment, for the detection of nightshade crop leaf disease. Their model enabled real-time disease diagnosis in the field, emphasizing the practical benefits of mobile-ready diagnostic tools [9]. Further advancements have been made by incorporating additional techniques into CNN. models to improve their performance. Thakur, Sheorey, and Ojha (2023) proposed VGGICNN, a lightweight CNN model for crop disease identification. Their model achieved high accuracy while maintaining computational efficiency, making it suitable for use in resource-constrained environments [10]. Similarly, Lu, Tan, and Jiang (2021) reviewed various CNN applications in plant leaf disease classification, providing valuable insights into the strengths and limitations of different CNN architectures and their potential for improving agricultural practices [11]. Rao et al. (2022) developed a deep bilinear CNN for plant disease classification, which demonstrated significant improvements in classification accuracy by leveraging bilinear pooling techniques. This approach highlighted the potential of advanced CNN architectures in achieving higher diagnostic precision [12]. Suresh, Gnanaprakash, and Santhiya (2019) analyzed the execution of different CNN architectures with various optimizations for the categorization of plant diseases. Their study provided a comprehensive evaluation of CNN models, identifying optimal configurations for enhancing model performance [13]. In addition to these advancements, researchers have explored resilient CNN architectures to improve robustness against variations in image data. Gokulnath and Usha Devi (2021) developed a resilient LF-CNN for identifying and classifying plant diseases. Their model showed significant improvements in handling diverse image datasets, ensuring reliable disease detection under varying conditions [14]. Sun et al. (2022) conducted study on the diagnosis of plant diseases using CNN, further validating the efficacy of deep learning models in accurately diagnosing plant diseases [15]. The integration of hybrid models has also been explored to enhance disease detection accuracy. Singh et al. (2022) proposed a hybrid feature-based disease detection system that combined CNNs with Bayesian Optimized SVM and Random Forest classifiers. Their approach achieved high accuracy in plant leaf disease detection, demonstrating the benefits of hybrid models in leveraging the strengths of multiple machine learning techniques [16]. Ma et al. (2023) developed a YOLOv5n algorithm incorporating attention mechanisms for maize leaf disease identification. Their model showed significant improvements in detection accuracy, highlighting the potential of incorporating attention mechanisms in CNN architectures [17]. Lastly, evolutionary feature optimization has been explored to enhance the performance of deep learning models in plant disease detection. Al-bayati and Ustünda (2020) developed an optimization of evolutionary features technique for plant leaves disease detection using deep neural networks. Their approach demonstrated significant improvements in model performance by optimizing feature selection processes [18]. In order to detect diseases in plant leaves, Geetharamani and Arun Pandian (2019) used a nine-layer CNN. By using deep learning techniques, they were able to achieve high classification accuracy [19]. These studies collectively underscore the transformative potential of deep learning and mobile technologies in plant disease detection. The integration of CNN models into mobile applications represents a promising direction for real-time agricultural disease management, addressing critical challenges faced by farmers worldwide. As the field continues to evolve, further research is essential to enhance the robustness and scalability of these models, ensuring their widespread adoption and impact in agriculture. Our model's superior performance can be attributed to several key factors: Advanced CNN Architecture, Extensive Dataset, Data Augmentation and Robust Training, Real-time Application.

### *Performance Comparison*

We compared our model's performance with several existing models to highlight its superior accuracy and robustness. Table I provides a summary of these comparisons. Our model achieved an accuracy of 98.14%, outperforming models such as the Bayesian Optimized SVM and Random Forest by Singh et al. (96.1%) and YOLOv5 + Swin Transformer by Ma et al. (95.2%). This significant improvement is attributed to our model's unique architecture and comprehensive dataset.

TABLE I: COMPARING ACCURACY & TECHNIQUE FROM RELATED WORKS.

| Paper | Technique | Accuracy |
|---|---|---|
| Singh et al. [16] | Bayesian Optimized SVM, and Random Forest | 96.1% |
| Li Ma et al. [17] | YOLOv5 + Swin Transformer | 95.2% |
| Arun Pandian et al. [19] | VGG16 | 92.87% |
| Mihir Kawatra et al. [20] | AlexNet with GAP Layer | 97.29% |
| **Our Model** | **CNN** | **98.14%** |

### *Problem Analysis*

According to the Food and Agriculture Organization (FAO), plant diseases account for approximately 40% of crop losses globally, translating to significant economic losses and threats to food security. For instance, fungal diseases alone are responsible for annual losses worth $60 billion worldwide. In a specific example, late blight affects about 16% of global potato production, causing severe economic damage. Bacterial spot in tomatoes can reduce yields by up to 90% under favorable conditions for disease development, highlighting the critical need for effective management strategies. Including statistical data on the prevalence and impact of these diseases can provide a clearer picture of the urgency and importance of developing robust plant disease detection systems.

## III. METHODOLOGY

This study endeavors to create a precise and resilient plant disease detection model utilizing convolutional neural networks (CNNs) and a comprehensive, varied dataset. The methodology encompasses pivotal stages, including dataset preparation, model design, training, evaluation, and incorporation into a mobile application for real-time implementation. The following sections detail each step of the process.

*A. Dataset Preparation*

*Data Collection*: A comprehensive dataset was sourced from Kaggle, comprising 87,867 images representing 26 different disease scenarios across 14 crops. The dataset comprises high-resolution images depicting both healthy and diseased plant leaves.

*Data Preprocessing*: To ensure the quality and consistency of the dataset, several preprocessing steps were undertaken:

- **Image Resizing**: All images were resized to a consistent dimension (256x256 pixels) to match the input size required by our CNN model.
- **Normalization**: Pixel values were normalized to a range of 0 to 1 to facilitate faster convergence during training.
- **Noise Removal**: Any noisy or irrelevant images were identified and removed to ensure the quality of the training data.
- **Data Augmentation**: To enhance dataset variability and mitigate overfitting, techniques such as rotation, flipping, and zooming were employed.

TABLE II: DATA DESCRIPTION

| ***Plant Names*** | **Leaf Types** | **Samples in Training** | **Samples in Validation** |
|---|---|---|---|
| ***Apple*** | Apple Scab | 2016 | 504 |
| | Black Rot | 1987 | 497 |
| | Cedar Apple Rust | 1760 | 440 |
| | Healthy | 2008 | 502 |
| ***Blueberry*** | Healthy | 1816 | 454 |
| ***Cherry*** | Powdery Mildew | 1683 | 421 |
| | Healthy | 1826 | 456 |
| ***Corn*** | Gray Leaf Spot | 1642 | 410 |
| | Common Rust | 1907 | 477 |
| | Northern Leaf Blight | 1908 | 477 |
| | Healthy | 1859 | 465 |
| ***Grape*** | Black Rot | 1888 | 472 |
| | Black Measles (Esca) | 1920 | 480 |
| | Leaf Blight | 1722 | 430 |
| | Healthy | 1692 | 423 |
| ***Orange*** | Citrus Greening | 2010 | 503 |
| ***Peach*** | Bacterial Spot | 1838 | 459 |
| | Healthy | 1728 | 432 |
| ***Bell Pepper*** | Bacterial Spot | 1913 | 478 |
| | Healthy | 1988 | 497 |
| ***Potato*** | Early Blight | 1939 | 485 |
| | Late Blight | 1939 | 485 |
| | Healthy | 1824 | 456 |
| ***Raspberry*** | Healthy | 1781 | 445 |
| ***Soybean*** | Heathy | 2022 | 505 |
| ***Squash*** | Powdery Mildew | 1736 | 434 |
| ***Strawberry*** | Leaf Scorch | 1774 | 444 |
| | Healthy | 1824 | 456 |
| ***Tomato*** | Bacterial Spot | 1702 | 425 |
| | Early Blight | 1920 | 480 |
| | Late blight | 1851 | 463 |
| | Leaf Mold | 1882 | 470 |
| | Septoria Leaf Spot | 1745 | 436 |
| | Spider Mites | 1741 | 435 |
| | Target Spot | 1827 | 457 |
| | Tomato Yellow Leaf Curl Virus | 1961 | 490 |
| | Tomato Mosaic Virus | 1790 | 448 |
| | Healthy | 1926 | 481 |

## *B. Model Design*

*CNN Architecture*: A novel CNN architecture was designed to optimize plant disease detection performance. The architecture includes:

- **Convolutional Layers**: Multiple convolutional layers were used to extract features from the images, followed by ReLU activation functions.
- **Pooling Layers**: Max-pooling layers help minimize the dimension of the feature maps while keeping critical features.
- **Fully Connected Layers**: Dense layers to combine features and enable classification.
- **Dropout Layers**: Dropout layers were utilized to counter overfitting by randomly deactivating a fraction of input units during training.

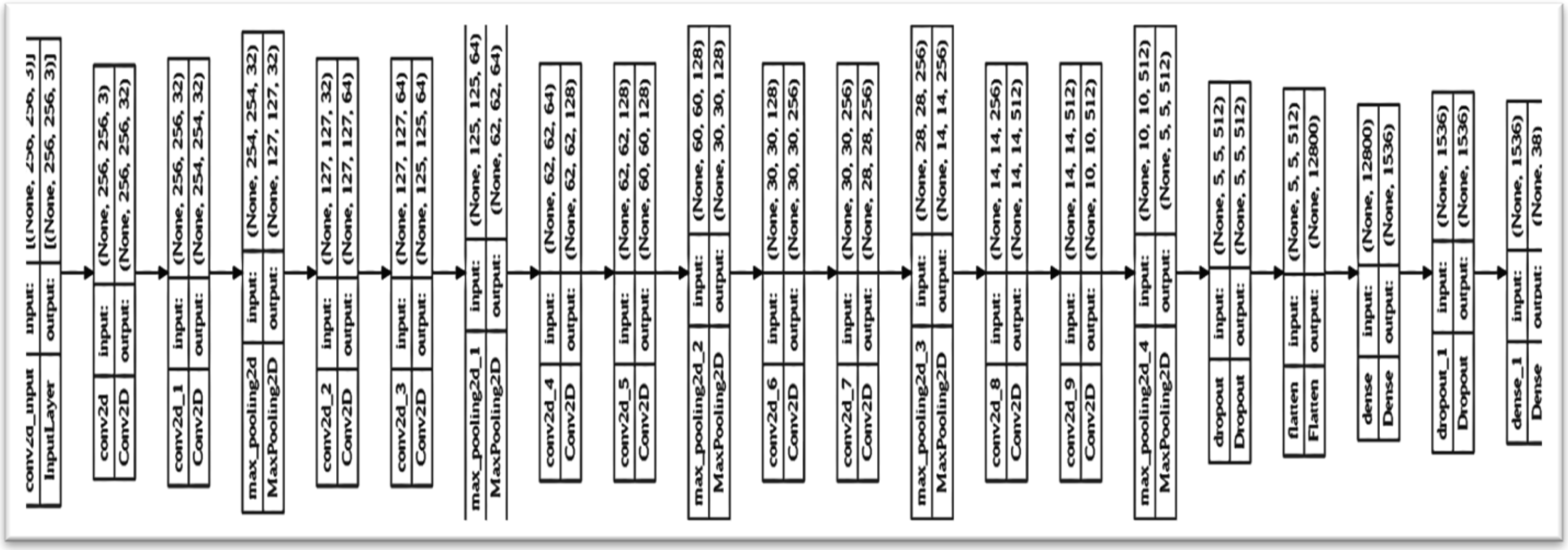

Figure 1: CNN Model Architecture

*Hyperparameters and Settings*: We carefully selected the hyperparameters for our model to optimize performance:

- **Learning Rate**: A learning rate of 0.0001 was selected to guarantee stable convergence.
- **Batch Size**: We used a batch size of 32, balancing computational efficiency and model accuracy.
- **Epochs**: Training the model for 15 epochs proved adequate to attain high accuracy without encountering overfitting.

*Data Splitting*: The dataset was partitioned into three subsets:

- **Training Set**: The model was trained using 75% of the images.
- **Validation Set**: During training, 25% of the images were allocated for validating the model's performance.
- **Test Set**: A separate folder containing 33 images was used exclusively for testing and hyperparameter tweaking.

*Statistical Analysis and Performance Evaluation:*

- **Proportion of Plants Affected:** We conducted a statistical analysis to determine the proportion of plants affected by each type of disease, providing context and significance to our study.
- **Performance Metrics:** Metrics such as accuracy, precision, recall, and F1-score were computed to evaluate the model's performance. A detailed comparison with existing models was included to highlight the improvements achieved with our approach.

*Mobile App Integration:*

- **App Development:** Details regarding the mobile app development, including the database and technology stack used, were provided. The app integrates seamlessly with the CNN model, allowing real-time disease detection and providing actionable insights to the users.

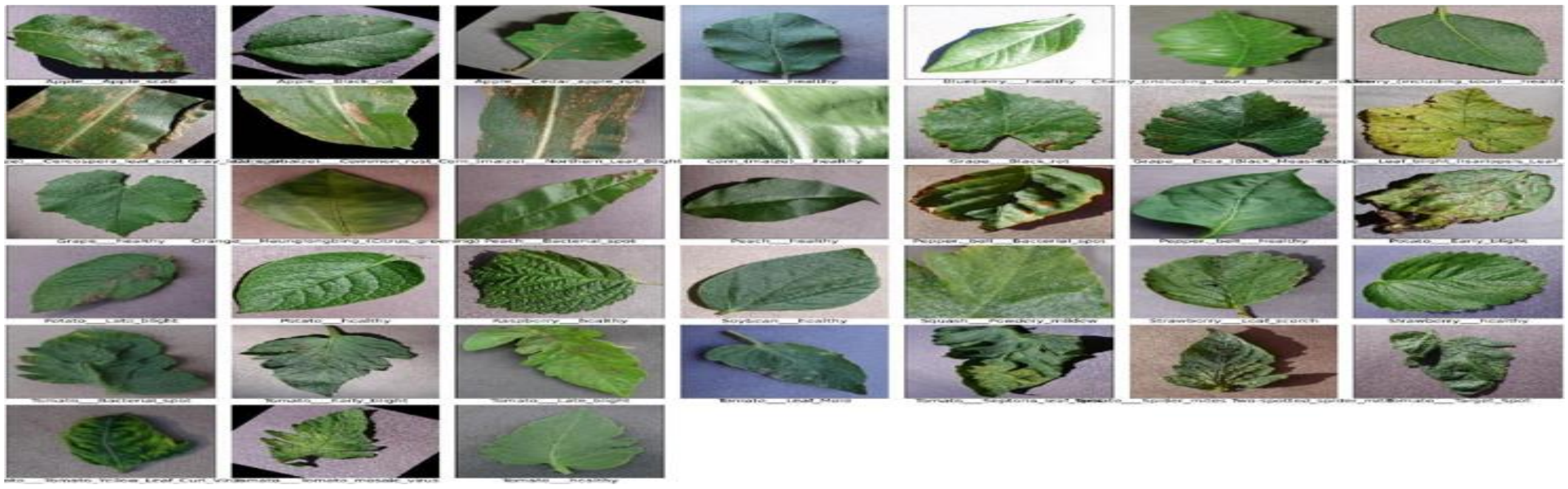

Figure 2: Sample Images from Each Class of Dataset

The model summary with different parameter values is presented in Table II and Table III.

TABLE III: MODEL OVERVIEW 1

| LAYER | OUTPUT | PARAMETER |
|---|---|---|
| conv2d | (256, 256, 32) | 896 |
| conv2d_1 | (254, 254, 32) | 9248 |
| max_pooling2d | (127, 127, 32) | 0 |
| conv2d_2 | (127, 127, 64) | 18496 |
| conv2d_3 | (125, 125, 64) | 36928 |
| max_pooling2d_1 | (62, 62, 64) | 0 |
| conv2d_4 | (62, 62, 128) | 73856 |
| conv2d_5 | (60, 60, 128) | 147584 |
| max_pooling2d_2 | (30, 30, 128) | 0 |
| conv2d_6 | (30, 30, 256) | 295168 |
| conv2d_7 | (28, 28, 256) | 590080 |
| max_pooling2d_3 | (14, 14, 256) | 0 |
| conv2d_8 | (14, 14, 512) | 3277312 |
| conv2d_9 | (10, 10, 512) | 6554112 |
| max_pooling2d_4 | (5, 5, 512) | 0 |
| dropout | (5, 5, 512) | 0 |
| flatten | (12800) | 0 |
| dense | (1536) | 19662336 |
| dropout_1 | (1536) | 0 |
| Dense_1 | (38) | 58406 |

TABLE IV: MODEL OVERVIEW 2

| Total_Parameters | 30724422 |
|---|---|
| Trainable_Parameters | 30724422 |
| Non-trainable Parameters | 0 |

***Evaluate Confusion Metrics***:

- The confusion matrix showed that the majority of the misclassifications were among similar disease classes, which can be challenging even for human experts.
- The model performed exceptionally well in distinguishing between distinct disease classes.

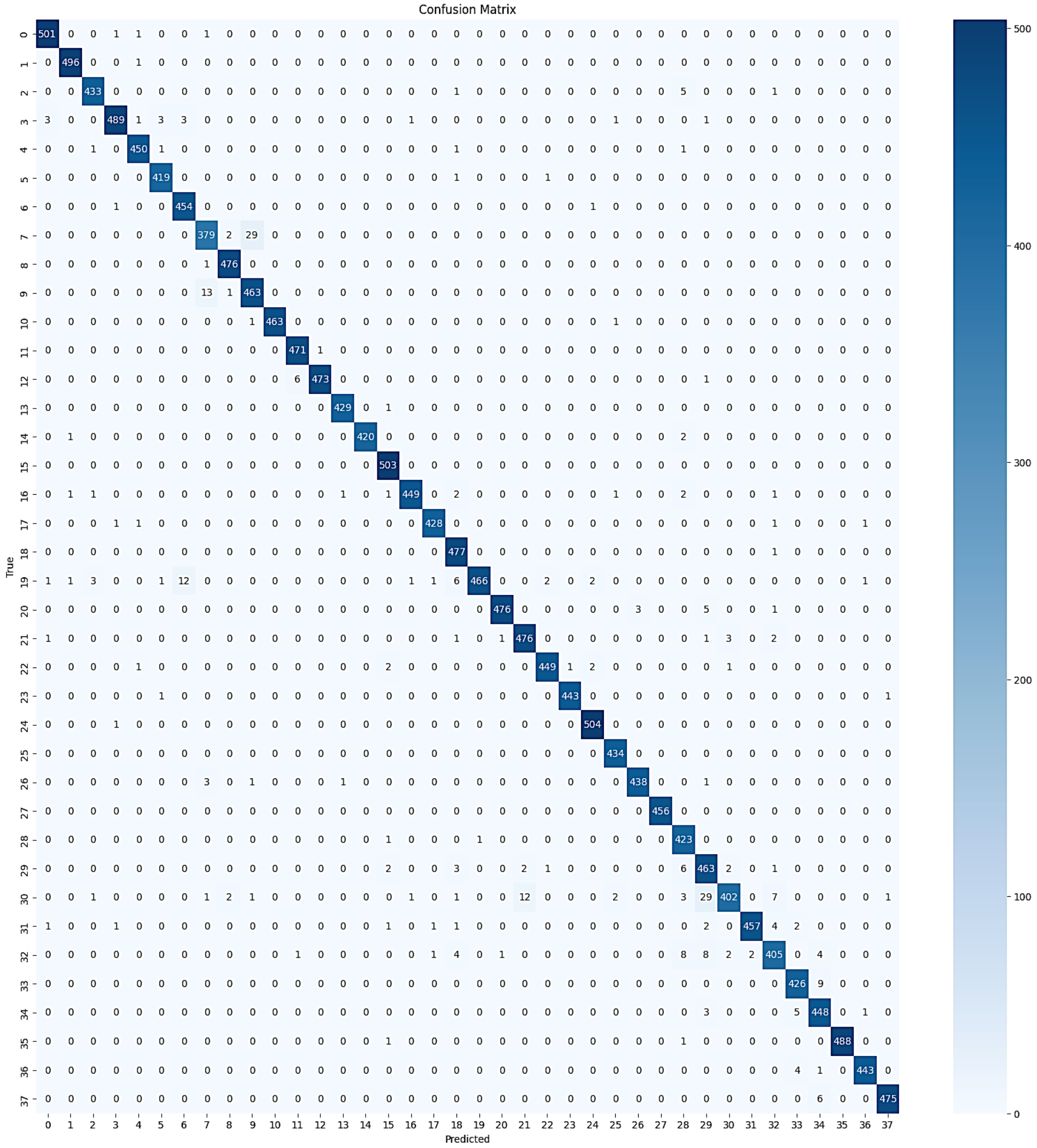

Figure 3: Confusion Matrix

## IV. EXPERIMENTAL RESULT

The experimental results of this study showcase the effectiveness of the suggested CNN architecture in precisely detecting and categorizing illnesses in plants. The model underwent training and evaluation using a comprehensive dataset comprising 87,867 images, encompassing 26 different disease scenarios across 14 crops. The dataset was split into training, validation, and test sets to guarantee comprehensive training and evaluation of the model's performance.

**Training and Validation Accuracy**:

After training the model, we achieved a final training accuracy of approximately 99.7% and a validation accuracy of 98.14%. These high accuracy rates indicate that the model has learned to classify the plant disease images effectively. The loss values were also low, further confirming the model's performance.

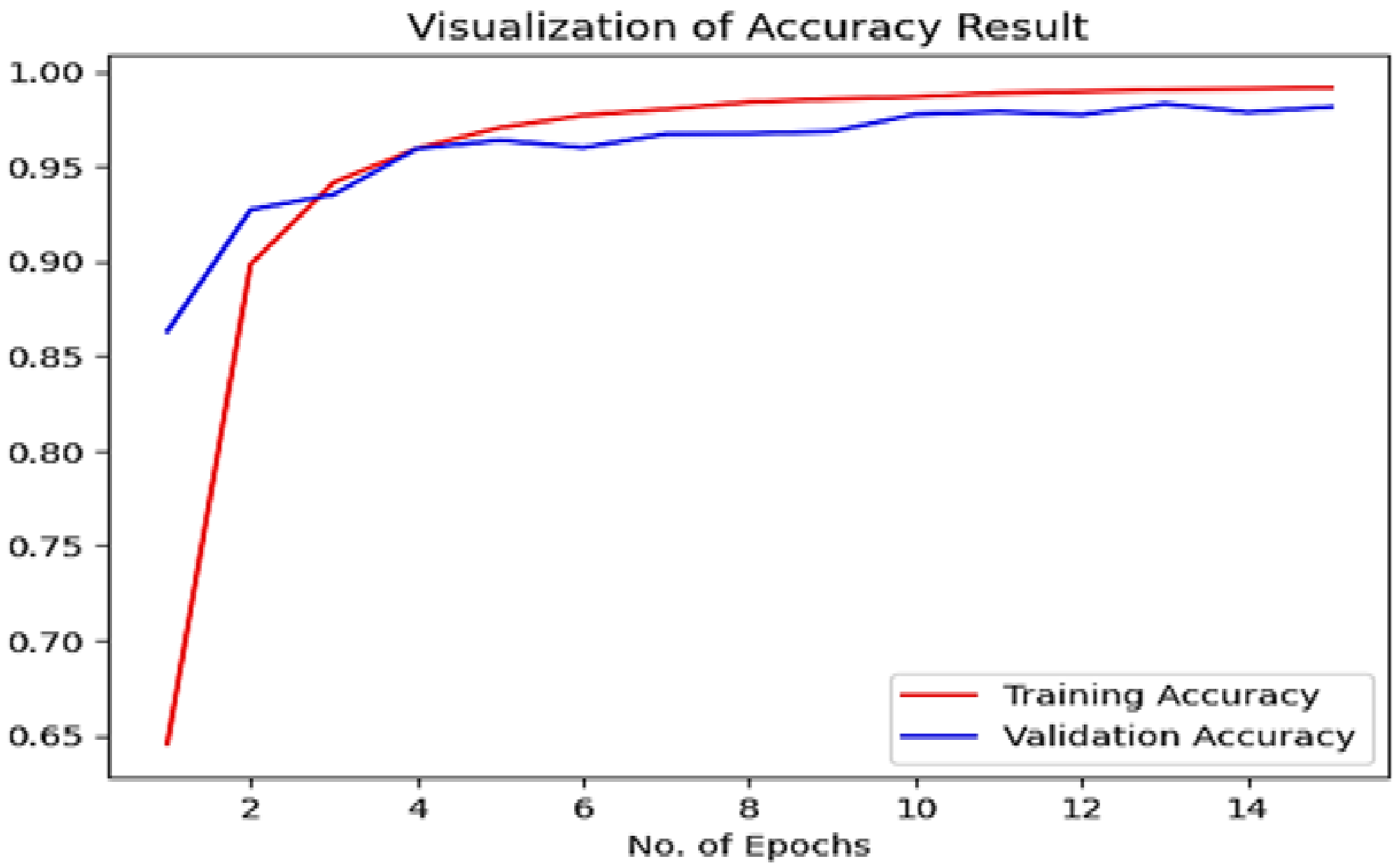

Figure 4: Accuracy Visualization Curve

**Loss Curves**:

- The convergence of the training and validation loss curves indicates that the model effectively learned from the training data and demonstrated good generalization to unseen data.
- The final training and validation losses were low, further indicating a well-trained model.

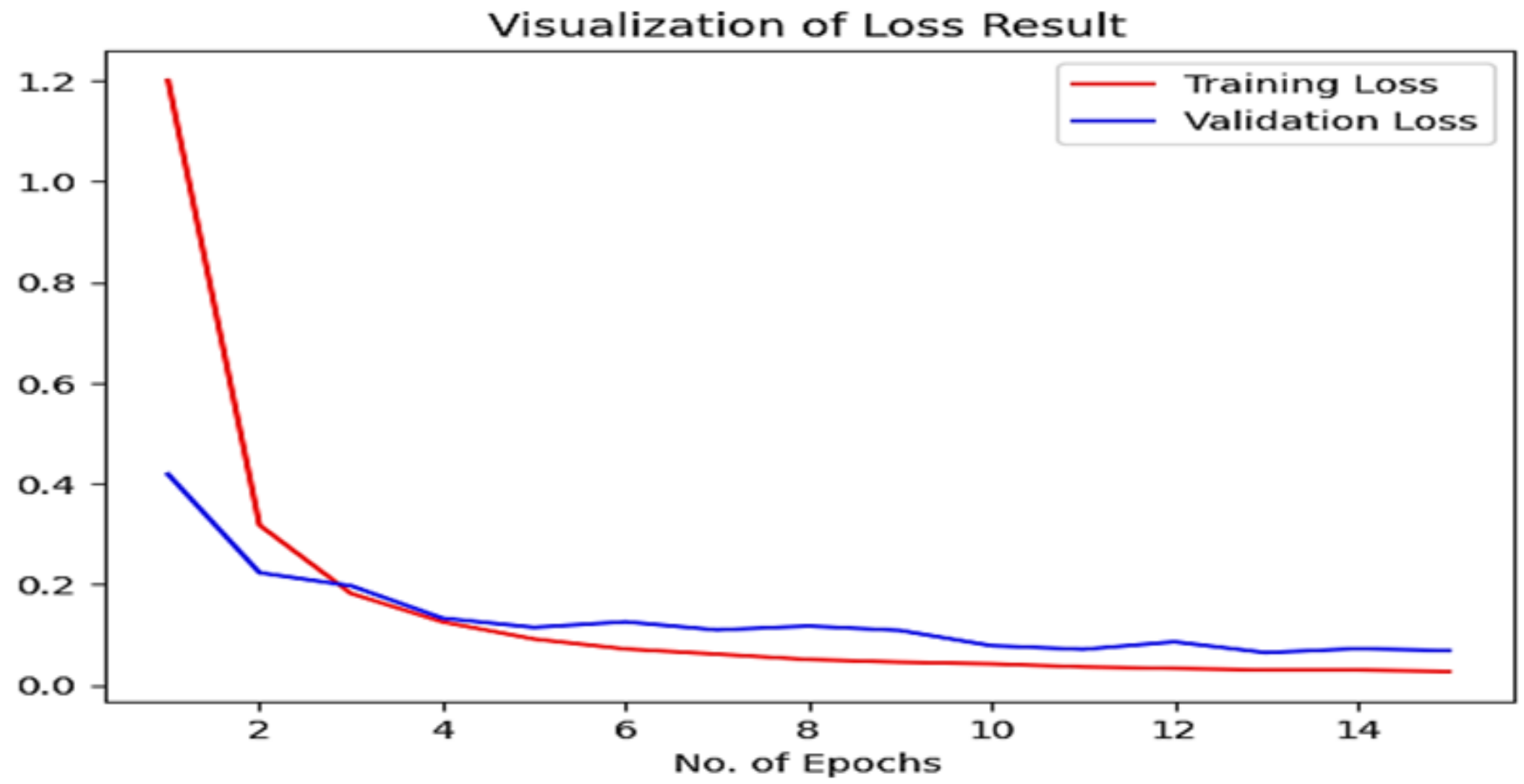

Figure 5:  Loss Visualization Curve

We present the performance metrics of our trained model after 15 epochs of training. The table provided below, referred to as TABLE IV, presents a summary of the training and validation accuracy, along with the corresponding loss values. These metrics offer insights into the efficacy of our model in classifying plant disease images.

TABLE V: TRAINING & VALIDATION DATA

| ***Epoch*** | ***Loss*** | ***Accuracy*** | ***Validation Loss*** | ***Validation Accuracy*** |
|---|---|---|---|---|
| *1st* | 1.2012 | 0.6456 | 0.4188 | 0.8631 |
| *2nd* | 0.3175 | 0.8983 | 0.2230 | 0.9272 |
| *3rd* | 0.1819 | 0.9415 | 0.1971 | 0.9352 |
| *4th* | 0.1254 | 0.9594 | 0.1327 | 0.9593 |
| *5th* | 0.0913 | 0.9704 | 0.1145 | 0.9637 |
| *6th* | 0.0716 | 0.9769 | 0.1255 | 0.9599 |
| *7th* | 0.0613 | 0.9801 | 0.1094 | 0.9670 |
| *8th* | 0.0506 | 0.9839 | 0.1172 | 0.9672 |
| *9th* | 0.0454 | 0.9855 | 0.1084 | 0.9684 |
| *10th* | 0.0420 | 0.9866 | 0.0782 | 0.9774 |
| *11th* | 0.0359 | 0.9885 | 0.0708 | 0.9788 |
| *12th* | 0.0329 | 0.9895 | 0.0857 | 0.9772 |
| *13th* | 0.0298 | 0.9904 | 0.0643 | 0.9829 |
| *14th* | 0.0298 | 0.9908 | 0.0722 | 0.9784 |
| ***15th*** | **0.0268** | **0.9914** | **0.0683** | **0.9814** |

Performance Analysis: One important indicator is accuracy, which expresses the percentage of cases properly identified relative to all instances. The suggested CNN model showed a higher degree of accuracy in classifying plant diseases with an accuracy of 98.14% on the test set.

TABLE VI: PRECISION RECALL & F1 SCORE

| | |
|---|---|
| Precision Score | 98.17 % |
| Recall Score | 98.14 % |
| F1 Score | 98.13 % |

### *Mobile App Development*

The disease detection model was seamlessly integrated into a user-friendly mobile application designed for Android devices, offering farmers and agricultural workers an intuitive tool for real-time plant disease diagnosis. The app, with a compact size of approximately 142 MB, is optimized for offline functionality, allowing users to diagnose plant health on-site without the need for an internet connection.

**User Interface and Experience**

The app's interface is designed with simplicity and efficiency in mind, ensuring ease of use for individuals with varying levels of technical expertise. Users can either upload images from their phone's gallery or use the device's camera to capture new images of plant leaves for analysis.

**Results are color-coded and visually enhanced for immediate interpretation:**

- **Green Healthy Status:** If the plant leaf is healthy, the app displays the result in **green** accompanied by a **green leaf emoji** 🌿. This instant visual cue reassures users of the plant's good health.
- **Red Diseased Status:** In cases where the model detects a disease, the result is highlighted in **red** with a **bacteria emoji** 🦠, signaling the need for prompt action. This color-coding helps users quickly identify issues that require attention.
- **Emoji Integration for Plant Identification:** To further enhance the user experience, the app shows the identified type of plant together with the corresponding emoji. Emojis for tomatoes '🍅', apples '🍏', corn '🌽' and so on. This playful yet effective usage of emojis not only makes the application visually appealing, but it also assures that the results are easily accessible for all farmers, whether they are literate, illiterate, or have difficulties with sight. By mixing emojis with color codes, it creates a joyful and inclusive experience that breaks down language barriers.

**Practical Features**

- **Offline Functionality:** The app operates fully offline, ensuring that farmers in remote areas with limited connectivity can still benefit from accurate disease diagnostics.

- **Lightweight and Efficient:** Instead of having a built-in storage facility, the app utilizes the phone's gallery or camera to fetch images. Users can either upload images directly from their gallery or use the device's camera to capture new images for analysis. This design choice minimizes the app's storage requirements while ensuring compatibility with a wide range of devices.

The integration of these visual and functional enhancements transforms the app into a powerful, accessible tool for modern agriculture, offering immediate, actionable insights into plant health.

**Real-Time Testing with Mobile Application**

To validate the model's practical applicability, it was integrated into a mobile app and tested in real-time scenarios. The goal was to develop an intuitive smartphone application that allows users to take or upload pictures and get detection results in real time. The following observations were made:

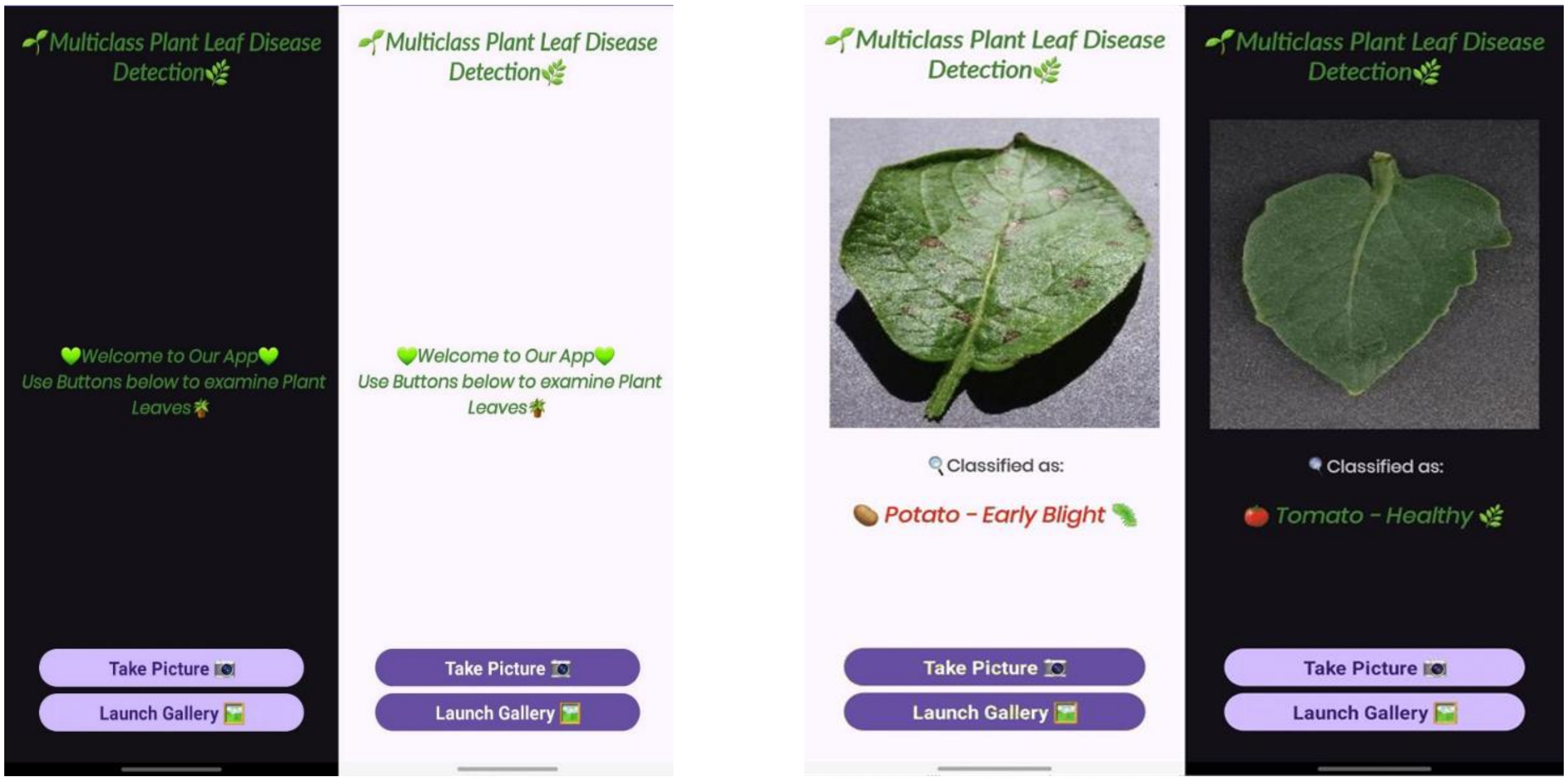

Figure 6: Home Page of the Android Application (in the left) and Disease Detection (in the right)

## V. CONCLUSION

The successful development and implementation of the proposed CNN model for plant disease detection represent a significant advancement in agricultural technology. By providing an accurate, reliable, and practical solution for identifying plant diseases, this research enhances the sustainability and productivity of global agriculture. The integration of advanced deep learning models into mobile applications opens new avenues for real-time agricultural disease management, empowering farmers with the tools needed to address critical challenges effectively. Future work will focus on expanding the dataset, refining the model, and exploring additional applications of this technology in other areas of agriculture. The app's offline functionality, enabled by a lightweight TensorFlow Lite model, allows real-time disease diagnosis even in remote areas with limited connectivity. Although the current implementation does not integrate directly with IoT devices, such as drones and smart sensors, the model's architecture is designed with scalability in mind. This makes future integration with IoT-based systems feasible, potentially enabling comprehensive, continuous monitoring of large agricultural fields. Such integration would allow for automated and proactive disease detection, enhancing the overall efficiency and effectiveness of agricultural management